# Hierarchical Composable Optimization of Web Pages

Ronen Barenboim [#1], Edward Bortnikov [*2], Nadav Golbandi [*3], Amit Kagian [*4], Liran Katzir [*5]
Ronny Lempel [*6], Hayim Makabee [*7], Scott Roy [#8], Oren Somekh [9]

[#]*Yahoo! Inc., Sunnyvale, CA, USA*
[1]`ronenbar@yahoo-inc.com`
[8]`sroy@yahoo-inc.com`

[*]*Yahoo! Labs, Matam, Haifa 31905, Israel*
[2]`ebortnik@yahoo-inc.com`
[3]`nadavg@yahoo-inc.com`
[4]`akagian@yahoo-inc.com`
[5]`lirank@yahoo-inc.com`
[6]`rlempel@yahoo-inc.com`
[7]`makabee@yahoo-inc.com`
[9]`orens@yahoo-inc.com`

*Abstract*—The process of creating modern Web media experiences is challenged by the need to adapt the content and presentation choices to dynamic real-time fluctuations of user interest across multiple audiences. We introduce FAME – a Framework for Agile Media Experiences – which addresses this scalability problem. FAME allows media creators to define abstract *page models* that are subsequently transformed into real experiences through algorithmic experimentation. FAME's page models are hierarchically composed of simple building blocks, mirroring the structure of most Web pages. They are resolved into concrete page instances by pluggable algorithms which optimize the pages for specific business goals. Our framework allows retrieving dynamic content from multiple sources, defining the experimentation's degrees of freedom, and constraining the algorithmic choices. It offers an effective separation of concerns in the media creation process, enabling multiple stakeholders with profoundly different skills to apply their crafts and perform their duties independently, composing and reusing each other's work in modular ways.

## I. INTRODUCTION

Digital media websites have mainly grown out of traditional media outlets, e.g. TV and printed press. Accordingly, such sites produce their pages in a process resembling traditional media, i.e. through an editorial board that decides on the various aspects of the page – its layout, which content items (including ads) to place where in the layout, and how to render (or format) the content. Editors of printed and broadcast media are typically senior journalists, who mostly apply human judgment, intuition and experience in creating their editions. However, two gating factors limit their ability to optimize, in a methodological manner, the product they put in front of their audience.

First, print editors have no way of tracking (other than through, perhaps, small focus groups) how their product was consumed and how it "performed", i.e. which stories, sections or ads resonated well with the readership/viewership and which didn't. Circulation and ratings trends take many weeks to develop, and are difficult to attribute to any specific decisions made by the editors. In contrast, through proper instrumentation, digital media editors can get immediate feedback on the consumption of their product. It is easy to track in real-time the number of times each story (or ad) was clicked, each video played, and each survey answered. Tracking unique visitors and repeat visitors in an accurate manner over any time period is also possible. Unlike their counterparts from the press, editors of online media have no lack of data – on the contrary, they perhaps have too much data to humanly reason about.

A second major difference between printed and broadcast media and online media revolves around the concept of an "edition". Printed media outputs editions at fixed intervals, e.g. daily or weekly, and prints a limited number of different versions per edition (e.g. editions may vary slightly between locales). Television programs are also produced at a regular pace, with minor variability. In contrast, online media isn't bound to the concept of editions, and content can rotate in and out of a Web page in real time. Lifetimes of content items vary, i.e. some items can remain served much longer than others. Furthermore, the number of different served versions of a page may, theoretically, equal the number of users to whom it was served.

Obviously, editorial attention does not scale to support the shift from a small discrete space of editions to a large continuum. Human analysis of complex instrumentation is also not scalable. With proper automation, however, online media can tap the ability to produce a continuum of editions to *experiment* with a huge variety of generated pages. The power of interaction instrumentation and the resulting usage statistics can then be leveraged to optimize the media experience - namely, choose the settings that produce the most favorable experience at every point in time.

In recent years, reinforcement learning has been applied in various contexts to optimize digital media in the sense described above, mainly by tapping click-through rates. Examples include optimizing search results ranking [12], [14] and main story optimization [1], [2]. Whereas the above works mainly revolved around optimizing content, other efforts leveraged user interactions to optimize presentation, e.g. layout of content on mobile devices [9]. These efforts demonstrated that algorithmic optimization of media can outperform editorial decisions and intuitions.

We envision that editors of future media sites will oversee their product at a high level, delegating many functions to algorithmic machinery, to keep up with scalability challenges. They will leverage experience and intuition to define the desired experiences in abstract terms, through *logical page models*. Rather than fully specifying the page, these models will include *degrees of freedom* on its layout, content and formatting, *constraints* on how those degrees of freedom may be jointly resolved, and the *target function* to pursue. For example, degrees of freedom may include specifying a large content pool from which several stories are to be selected, choosing from among several layouts, and choosing formatting or algorithmic configuration parameters. Constraints may be imposed on the joint resolutions of the degrees of freedom, thus introducing dependencies between those resolutions. Target functions may factor in user engagement (click-through rates, repeat visits, etc.) as well as monetization aspects. The actual optimization of the page, within the parameters defined by the editors, will be algorithmic.

This paper introduces FAME – a Framework for Agile Media Experiences. We present the vision and prototype of the system, parts of which are already in production at a major Internet portal. FAME defines a hierarchical logical model for describing complex self-optimizing web pages. This model allows a fine-grained interplay between algorithmic decisions and editorial control. FAME allows independent plug-ins to optimize various decisions on the page, and orchestrates those plug-ins so that their joint output will satisfy the constraints while performing well in the target function. Essentially, the FAME execution engine explores the space of possible page instantiations, attempting to converge to the best performing one.

Our architecture emphasizes the decoupling and composability of all artifacts – dynamic data sources, degrees of freedom, constraints and optimization plug-ins – across multiple pages. Thus, the different stakeholders that participate in the media creation process can perform their roles in a mostly independent and repeatable fashion, while utilizing profoundly different skill sets. To the best of our knowledge, no existing page optimization system achieves this level of separation-of-concerns, while capturing the complex structure of modern Web pages as well as accommodating state-of-the-art optimization algorithms.

The rest of this paper is organized as follows. Section II describes the ecosystem of online media creation, and explains how the the FAME framework combines the efforts of the various stakeholders in agile creation of optimized media pages. Section III surveys related work, and emphasizes our contributions. Section IV defines our logical page model, and specifies how degrees of freedom for experimentation are injected into the pages. Section V extends the page model by adding constraints that enforce editorial validity of the result. Section VI introduces the model execution engine. The FAME machinery applies algorithmic plug-ins to optimize the actual content and layout choices within the aforementioned degrees of of freedom, while adhering to the user-defined constraints. Section VII describes FAME's infrastructure for collecting and harnessing real-time user feedback, and Section VIII concludes the paper.

## II. AGILE MEDIA CREATION

The process of creating new media experiences revolves around two resource pools: the raw *content* pool (articles, images, video clips, blog posts, tweets, ads, etc.), and the much smaller *user experience design (UED)* pool (web pages' layouts, navigational bars, functional widgets, image carousels, etc.). The first pool comes from a variety of sources (news and social feeds, multimedia repositories, etc.), whereas the second one is typically created by the media product's *UED specialists*. The Web page ultimately served to the user is optimized for multiple goals, usually striking some balance between user satisfaction and business objectives.

FAME is a framework for systematic experimentation with combinations of content and design while optimizing for the desired goals. It offers a separation of concerns that allows people with very different skill sets to address different aspects of media experience creation independently. The three pillars around which these aspects can be depicted are shown in Figure 1:

**Degrees-of-Freedom Management**. A *product owner* in charge of a particular experience creates a *logical page model* that defines the eventual impression, subject to the desired level of experimentation. For example, a media site's front page owner may use a fixed layout, allocating one slot to an ad, and two other slots to any combination of news, sports and entertainment items. Alternatively, one might use a fixed set of data sources, and experiment with two possible layouts – e.g., varying the ads' locations.

**Constraint Definition**. *Media editors* responsible for maintaining semantic consistency of an experience may cast a set of constraints that mandate (or conversely, rule out) certain content and graphic design combinations. For example, a website's content editor may require that only hard-core news items become the centerpiece of its main page, or demand that no two rich media spots are rendered next to each other. Moreover, the editor may enforce the same constraints uniformly across all pages.

**Content and Business Optimization**. *Optimization experts* tackle the problems of optimizing the media and layout selection subject to various goals (user interaction, monetization, etc), which are typically set by *business owners*. For example, one team may focus on the problem of choosing and ranking stories to populate a list of story links, while a different team may solve an ad-slotting problem. Their deliverables are software building blocks that can be composed on demand in optimizing experiences. Eventually, a relatively small toolkit of optimization algorithms will be applicable to be reused in many different media experiences that differ by layouts, content pools, target functions, etc.

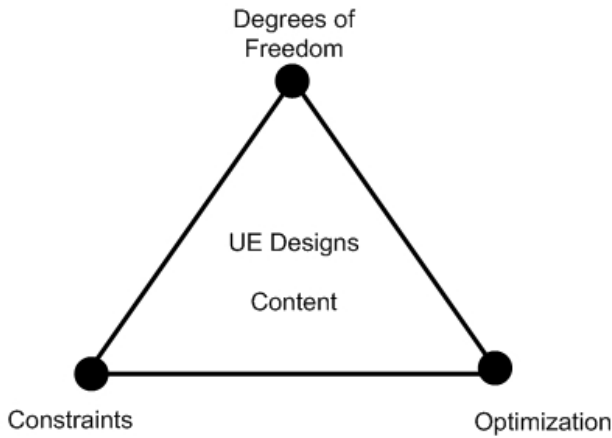

Fig. 1. **The FAME experimentation and optimization ecosystem.**

Manifesting the aspects above in separate, composable and reusable software and configuration artifacts allows isolating the business processes behind them, resulting in the agile rollout of optimized experiences. The framework allows all players above to virtually act together for a particular page impression, where in practice the various players may have never met, never tailored their decisions to the common context, and don't even know who and what their artifacts are interacting with. Thus, product owners are able to embed degrees of freedom in their pages – and algorithms that reconcile them – without having to engage with optimization experts. The algorithms that are used in the page may have never been used together before. The editorial policies can be enforced centrally, and independently of ongoing evolution in UED. Thus, new pages can be optimized by reusing and composing building blocks of previous optimization efforts, without the need of having all stakeholders actually meet and discuss the specifics of any given experience.

This agility comes at a price – inevitably, a tailored optimization process for each media experience might lead to a better result than an automatic integration of generic components. However, tailored approaches do not scale with the explosive number and rate of pages served by present-day sites. FAME's modular approach, which follows the best practices in architecting complex software systems [6], provides a solid methodology for this scalability problem.

III. RELATED WORK

Over a decade ago, Etzioni and Perkowitz coined the term *Adaptive Web Sites* to denote sites that automatically improve their organization and presentation by observing usage patterns [13]. Their vision is mostly realized today, as a significant amount of research has been invested since in improving Web experiences in light of metrics derived from usage patterns, most notably click through rates (CTR). The literature covers works that iteratively improve upon their metrics given usage feedback, as well as testing methodologies whereby a property experiments with multiple parameters (display options, algorithm versions, etc.) in parallel so as to choose a well performing setting.

Fiala [5] suggested a concern-oriented framework for adaptive web applications. The centerpiece of that work is a language for modular web document description, which features rules for adapting to user profiles. However, it does not deal with dynamic experimentation or constraint validation.

Kohavi et al. [11] published a survey on controlled experiments on the Web, ranging from A/B testing to multi-variable tests. One of their conclusions is that having the appropriate infrastructure for testing is key to achieving speedy and agile innovation. They note the importance of being able to test many ideas quickly, and to have the unsuccessful ones "fail fast". In a follow-up by some of the same authors [4], they focus on Web-specific pitfalls one should avoid when performing such experiments, giving many real-life examples.

Tang et al. [18] describe Google's framework for scaling the experimentation capacity of their Search site. They describe how multiple parameters are partitioned into *layers*, where experiments can modify one parameter per layer per *domain* (segment of traffic). The manually partitioned layers guarantee that parameters whose values may conflict with each other are never simultaneously tweaked.

In a deviation from standard A/B testing methodology, Radlinski et al. [14] compare search algorithms by interpreting CTR on an interleaved list of the results returned from both algorithms. The interleaving of results means each user is exposed to some amount of top results of both algorithms, unlike traditional A/B testing where each user is presented with the result set of just one algorithm. Moon et al. [12] suggest utilizing user feedback in the form of clicks for online learning tasks, such as re-ranking of top search results in temporally sensitive queries.

Many reinforcement learning works in recent years applied the well-known *Multi-Armed Bandit* problem [8] in optimizing Web experiences. Agarwal et al. [2], [1] discuss explore-exploit algorithms that tap user clicks on stories to optimize content on *Yahoo.com*. Their approach is a natural candidate to be plugged into our framework. Chakrabarti et al. [3] apply optimization to items (e.g. ads) whose performance degrades over time, therefore requiring constant vigilance (manifested in increased exploration) to ensure sustained performance. A different approach to a similar problem of time-sensitive ambiguous search results was applied by Syed et al. in [17]. They tackled queries whose prevalent meaning varies in time, and used an event classifier to essentially reset the bandit algorithm once they detect an event that causes the prevalent interpretation of the query to change.

In addition to the papers above, several commercial systems - both proprietary and open source - enable Web masters to easily set up multivariate experiments on their sites and either track metrics or automatically tune the sites' performance.

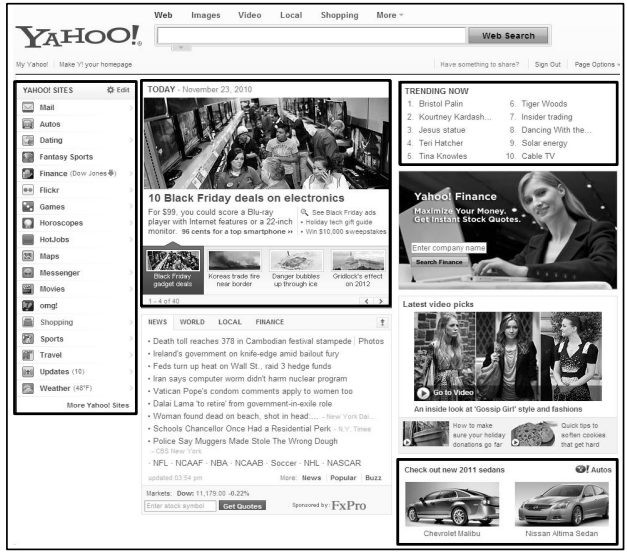 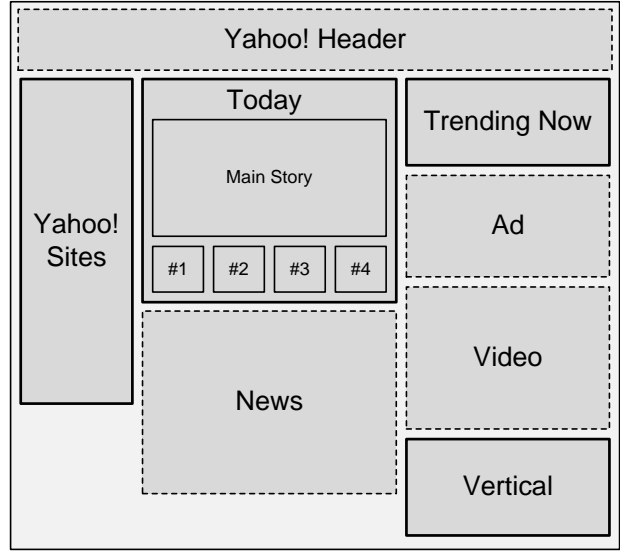

Fig. 2. The *Yahoo!* front page. (a) Physical representation; (b) Schematic representation.

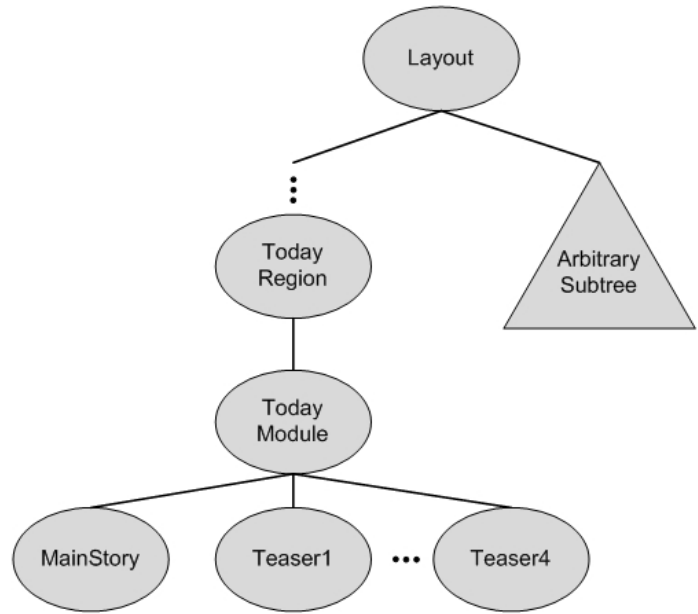

Fig. 3. The *Today* module's portion of the *logical page description*. (a) An XML representation. (b) An equivalent tree representation.

The Genetify open-source project[1] uses genetic algorithms to optimize Web pages. Web masters define groups of elements, where within each group, the various elements are display alternatives. Groups can be defined for HTML tags, CSS and Javascript. Various rewards can be associated with clicks on the outcome, and a genetic algorithm [10] is applied to explore user interactions across multiple versions of the page, converging (hopefully) to the instance of maximum reward. Some of Genetify's developers later applied similar technologies at SnapAds[2], a product for optimizing the performance of display ads. Advertisers create variations on elements of their ad, from which the software generates the cross-product of those variations as a space of candidate ads. Those candidates are pruned manually by the advertiser, and a genetic algorithm searches for the allowed instance that generates the best CTR.

[1] https://github.com/gregdingle/genetify/wiki/

Maxymiser[3] is a company that applies multivariate testing for content optimization and Web site personalization. Web masters define multiple variants of disjoint content areas, and the system explores the cross-product space of those variants to find the best-performing combinations. User segmentation is also applied, to discover different versions of the site that appeal to different populations.

Autonomy's Optimost service[4] is called by Javascript from the customers' sites. It applies multivariate testing for content, layout and presentation optimization of landing pages as well as ads. In addition to the experience improving based on usage, they provide extensive dashboards and analytic tools that allow Web masters to understand the ripple effect of changes in one page on the funnel of downstream pages users flow to within the site. User segmentation is also accounted for.

The Google Website Optimizer[5] uses Javascript injection to perform multivariate testing on pages. Success is tracked by the fraction of users who interact with the test page and ultimately visit a (typically different) *conversion page*.

Adobe's Test & Target[6] product suite offers multivariate testing, user segmentation and personalization at the single-user level, while having convenient integration with Adobe's user session tracking and logging product (SiteCatalyst).

Both Verster[7] and Webtrends Optimize[8] disclose using full as well as fractional factorial test designs to optimize campaign landing pages.

FAME improves upon the surveyed systems above in several aspects:

- Most aforementioned systems enable defining several sets of disjoint and independent options on the page, and running tests over the cross-product space where each page instance selects one option per set. Hence, the entire cross-product space is eligible for serving. As observed in [18], serving the entire cross-product space is a serious limitation due to unwanted interactions between combinations of options. Whereas [18] introduced the concept of layers, FAME uses a mechanism of constraints (see Section V) that avoids serving page instances with conflicting attributes.
- The algorithms that search the cross-product space and perform the optimization in the systems above are monolithic, proprietary and non-extensible. In contrast, FAME is an open platform that defines interfaces and extension points that allow multiple optimization algorithms (written by various parties) to be plugged in, each "resolving" specific portions of the page (see Section VI).
- The optimization processes in the surveyed systems do not leverage the hierarchical nature of Web pages. For example, they cannot readily postpone the resolution of degrees of freedom until content has been retrieved and/or other certain decisions have been made. FAME, in contrast, is designed for hierarchical optimization of the page (see Sections IV and VI).

IV. LOGICAL PAGE MODEL

It is convenient to describe complex web pages in hierarchical modular fashion. A nontrivial page is typically partitioned into a set of rectangular regions, in which content presentation entities that we call *modules* are embedded. For example, the snapshot of the Yahoo! front page depicted in Figure 2(a) is composed of eight regions, on top of which a header module and seven body modules are laid out. Focusing on the body's black-framed regions (see Figure 2(b)), the embedded modules are *Yahoo! sites* on the left, *Today* as the centerpiece, *Trending Now* at the top-right, and a *Yahoo! Vertical* module (currently showing content from *Yahoo! Autos*) at the bottom right corner. The hierarchical nature of the page is further manifested by the *Today* module, which includes five sub-regions, each depicting a news item with an image and some textual description.

Consider a concise representation of the *Today* module portion in XML format that natively captures the hierarchy (Figure 3(a)). This *logical page description* is genuinely separate from the *physical* aspects of Web pages (namely, the exact rendering and precise placement of the page elements) which constitute the final outcome (or product) of Web media servers. We assume that the XML contains enough information for the logical-to-physical transformation to occur. For example, in Figure 3(a), the logical page description (1) indicates that the *MainPage* physical layout must be applied, (2) associates the *Today* module with a region label, and (3) instructs to use the *Design7* design to render the experience. As this example is for illustrative purposes only, we do not concretely define the semantics of the tags, instead relying on their names and label attributes to convey their meaning. For ease of presentation, from this point on we forego the explicit XML representation, and instead exemplify our pages using the equivalent tree representation (nodes are XML tags, and edges correspond to direct nesting of one tag in another), as depicted in Figure 3(b).

The machinery which transforms logical page descriptions into actual browser-ready Web pages is built into *front-end* systems and is beyond the scope of this paper. For our purposes, it suffices to consider that a front-end system will receive users' HTTP requests for page generation, and will call FAME - a *back-end* system - to produce the logical page descriptions for responding to those requests. Logical page descriptions in XML are therefore FAME's output, for front-end consumption.

---

[2]http://www.snapads.com/
[3]http://www.maxymiser.com/
[4]http://promote.autonomy.com/
[5]https://www.google.com/analytics/siteopt/splash
[6]http://www.omniture.com/en/products/conversion/testandtarget
[7]http://www.vertster.com/
[8]http://www.webtrends.com/Products/Optimize

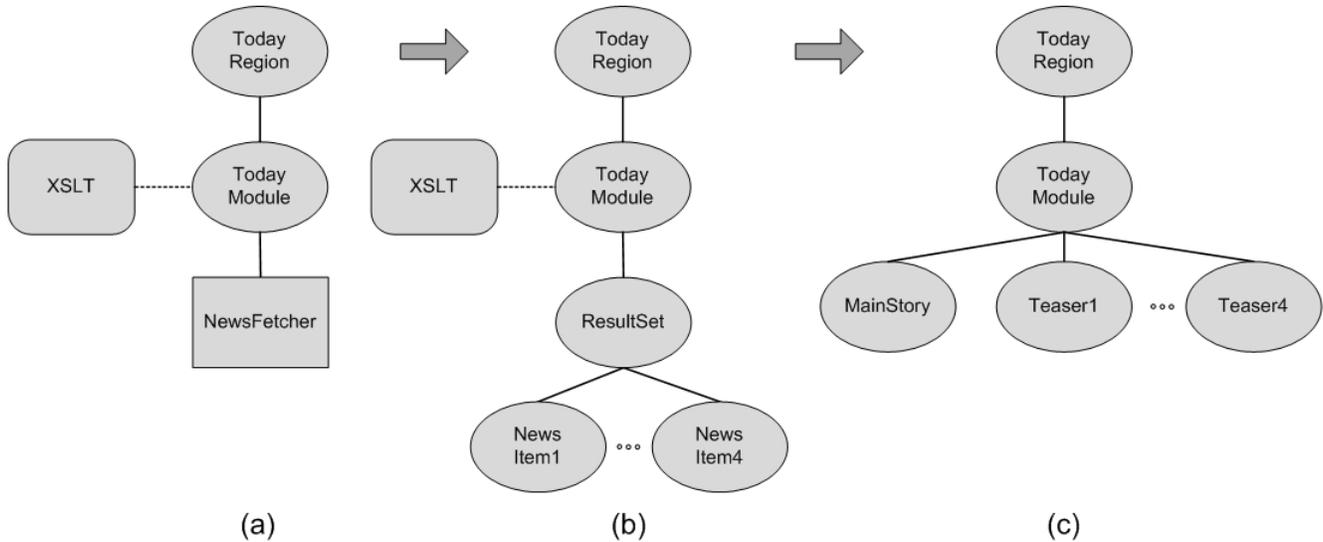

Fig. 4. Fetching dynamic content. (a) Fetcher; (b) Fetched content; and (c) XSLT transformation.

FAME's input, as passed from the front-end system, is a more abstract *logical page model*, also in XML. Logical page models capture the ecosystem discussed in Section II. They allow experimentation with degrees of freedom, optimization for user experience and business goals, and validation of editorial constraints. Syntactically, a logical page model is a logical page description into which platform-specific FAME tags are embedded. Each of these tags, which we call *operators*, has well-defined semantics and is associated with a unit of execution. That unit of execution, with the exact business logic it encapsulates for a specific operator instance, is pluggable into the model.

During execution (see Section VI), the logical page model (FAME's input) will gradually morph back into a logical page description (FAME's output) by a series of transformations defined by the operators. Concretely, the scope of an operator is the set of its subtrees (this reflects the hierarchical nature of the page composition process). Each operator evaluates to an XML tree free of further FAME constructs, which in turn is accessible to its parent operator(s). An operator can produce an *empty* result if it cannot be evaluated for some reason (e.g. there is no content for display, or there is no legitimate way to render two modules). Handling this situation is the consuming operator's responsibility – the latter can either rectify the problem, or itself return an empty result.

In what follows, we describe the FAME operator classes and their high-level semantics.

### A. Fetching Dynamic Content

Consider again the page depicted in Figure 2. Some of its elements are static (most notably, the header), while the others – e.g., the news stories in the *Today* module and the topics in *Trending Now* module – change dynamically in real time. In order to accommodate for such dynamic elements, FAME provides a *fetch* operator that fetches at runtime a certain number of elements from some content source. The content sources may be databases, content management systems, search engines, or any other service that is queryable at runtime. The operator evaluates to an XML subtree that encodes the *result set*, which in turn may be processed by some consuming operator. In order to compose consuming operators over result sets from specific sources, FAME allows XSLT plug-ins that perform schema transformations. For example, Figure 4 depicts the *Today* module as invoking a news fetcher, which at page generation time populates the logical page description with the actual news stories. XSLT then transforms the result set of news stories into the *Today* schema.

### B. Optimization with Degrees of Freedom

One of the pillars of FAME is the ability to declaratively incorporate degrees of freedom into the logical page description, thereby describing not a single page but rather a *space* of potential page instances. The *resolver* operators operate on multiple child subtrees, and produce a result that optimizes, exactly or approximately, some target function (user engagement, monetization etc.). Syntactically, they can be embedded at any point of the tree. The framework supports two types of such multivariate resolvers – *choice* and *map*.

The *choice* operator encompasses multiple alternative subtrees, and selects at runtime to instantiate one of them, effectively pruning the other alternatives from the document. Choice operators may be applied to user-visible elements of the page (e.g., populate region $X$ with one of modules $A$, $B$ or $C$) and also to configuration options (e.g. govern the ranking logic by parameter sets $D$ or $E$). Figure 5 shows the sub-tree corresponding to the *Yahoo! Vertical* region, with a degree of freedom allowing that region to be populated by an *Autos*, *Real Estate* or *Travel* module, and how that sub-tree is resolved when *Autos* is chosen.

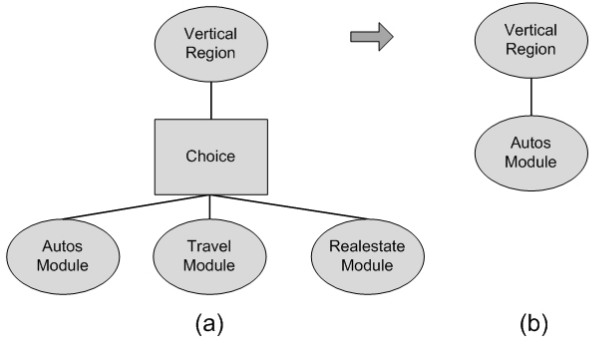

Fig. 5. Choice - Selecting one out of three Yahoo! verticals (a) Choice operator and; (b) Resolved tree.

The *map* operator is a generalization of *choice*, enabling to map $k$ out of $n$ $(n \geq k)$ *items* to $k$ *positions*. The main use of map is to test permutations of page elements – permute modules within regions, or items within a module. Building on the example from Figure 4, the news items fetched for the *Today* module may not necessarily be fetched in optimal order, and the number of potential stories fetched may certainly exceed four. Figure 6(a) shows how four placeholders are defined to act as map positions (for the four teaser stories), with the fetcher also embedded in the map. At runtime, the fetcher is executed (Figure 6(b)), its result set is transformed (by the XSLT associated with the map) into map items (Figure 6(c)), and the operator maps four of them into the placeholders, resulting in Figure 6(d). At that point, the situation is similar to what was depicted in Figure 4(b), thus completing the operator's evaluation. Note that the *Today* module XSLT is placing News item 1 also in the Main Story region (see Figure 2(a)).

The combination of the map and choice operators exemplified above describes a space of $3\binom{n}{4}$ pages, where $n$ is the number of news stories fetched. In practice (see Section VI-A), each resolver typically only considers a small number of viable outcomes per each user page request, making its computational overhead practically fixed. The entire space of possible pages is explored over many page requests over time.

We reiterate that resolver operators can be inserted anywhere in the logical page model. Accordingly, degrees of freedom can be defined over any aspect of the page: its content (*What* is shown), rendering format (*How* is it shown), and layout (*Where* is it shown). Furthermore, resolvers may be composed by nesting one operator in another. The example above showed a fetcher nested in a map, and similar compositions of choice and map operators are allowed.

## V. CONSTRAINING PAGE MODELS

Product owners and editors may wish to impose constraints on the space of page instances, to ensure that only pages that satisfy them will be generated. Such constraints are particularly useful in the following cases:

1) **Inter-operator constraint:** Certain combinations of choices and mappings may not mesh well together. These combinations may be deterministic (e.g. "never choose X in this operator if Y was chosen by that operator"), or may depend on the dynamics of the fetched content (e.g. to ensure deduplication or diversity of content). For instance, consider the predicate "The *Trending Now* module may not show a trend which appears in the headline of the *Today* module's main story".

2) **Intra-operator constraint:** Some attributes of the dynamic content may disqualify choices that are valid a-priori. For example, consider a constraint like "no more than 2 sports news items can appear in the *Today* module".

3) **User constraint:** Some choices may be undesirable for certain individuals. For example, items that a user has already consumed – or has repeatedly chosen to not consume – may no longer be valid for that user.

To support such a restriction mechanism, FAME includes a *constraints* operator. Any number of individual constraint predicates may be embedded within this operator. The overall semantics of *constraints* are that it returns an XML subtree for which all predicates are true. If no valid subtree can be found, the logical page model is not satisfied, and an empty result is generated[9]. Note that the semantics of *constraints* are purely declarative – e.g., an inter-operator constraint (see example above) does not imply order on the child operator's evaluation.

To make the expressiveness of constraints as general as possible while avoiding the definition of a proprietary syntax, the constraint predicates are functions written in a high-level programming language (in our implementation, JavaScript). A constraint function is a boolean method that returns *true* if the constraint is satisfied and *false* otherwise. Each constraint function has access (e.g., through XPath expressions) to the XML subtree rooted at the *constraints* tag. Thus, each constraint can access all elements of the subtree, and validate the predicate it represents.

Figure 7 depicts a partial tree representation of the Yahoo! front page, with intra- and inter-operator *constraints* operators marked in bold. Note how the inter-operator page-level constraint spans over the subtrees of both the *Today* and *Trending Now* modules, where each module includes a *map* resolver.

Constraints, like any other operator, can be inserted into the logical page model at any point of the XML tree. For performance reasons that will be detailed in Section VI, the best practice is to insert them at the deepest portion of the tree possible, i.e. as the least common ancestor of all the elements that any of the constraint functions refer to. In other words, at the immediate scope to which they apply. This will allow to validate decisions of nested operators – and to resolve any conflicts – as quickly as possible.

---

[9]Handling empty results is up to the parent operators – see Section IV.

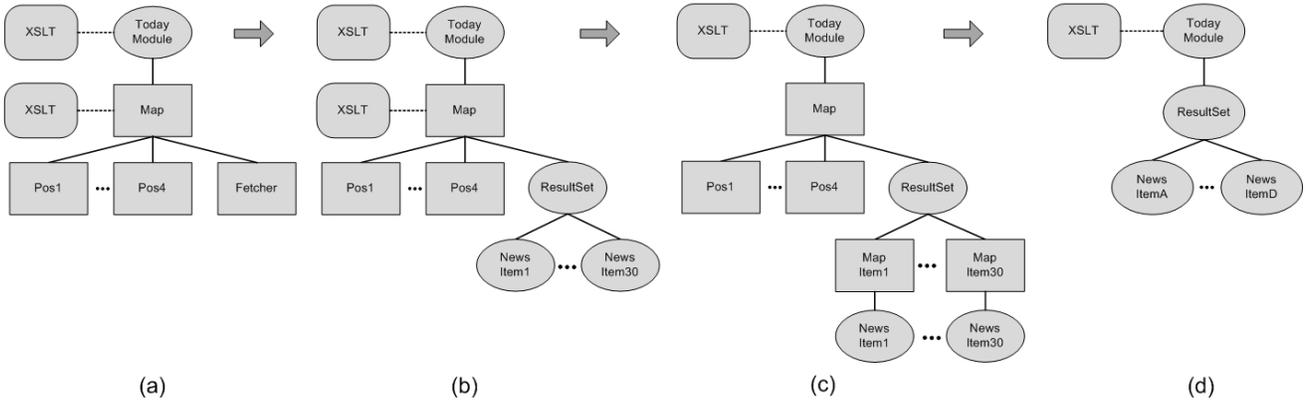

Fig. 6. Map - Mapping four out of thirty news items to four positions. (a) Fetcher; (b) Fetching content; (c) Map XSLT transformation; and (d) News mapping (for XSLT transformation see Figure 4(c)).

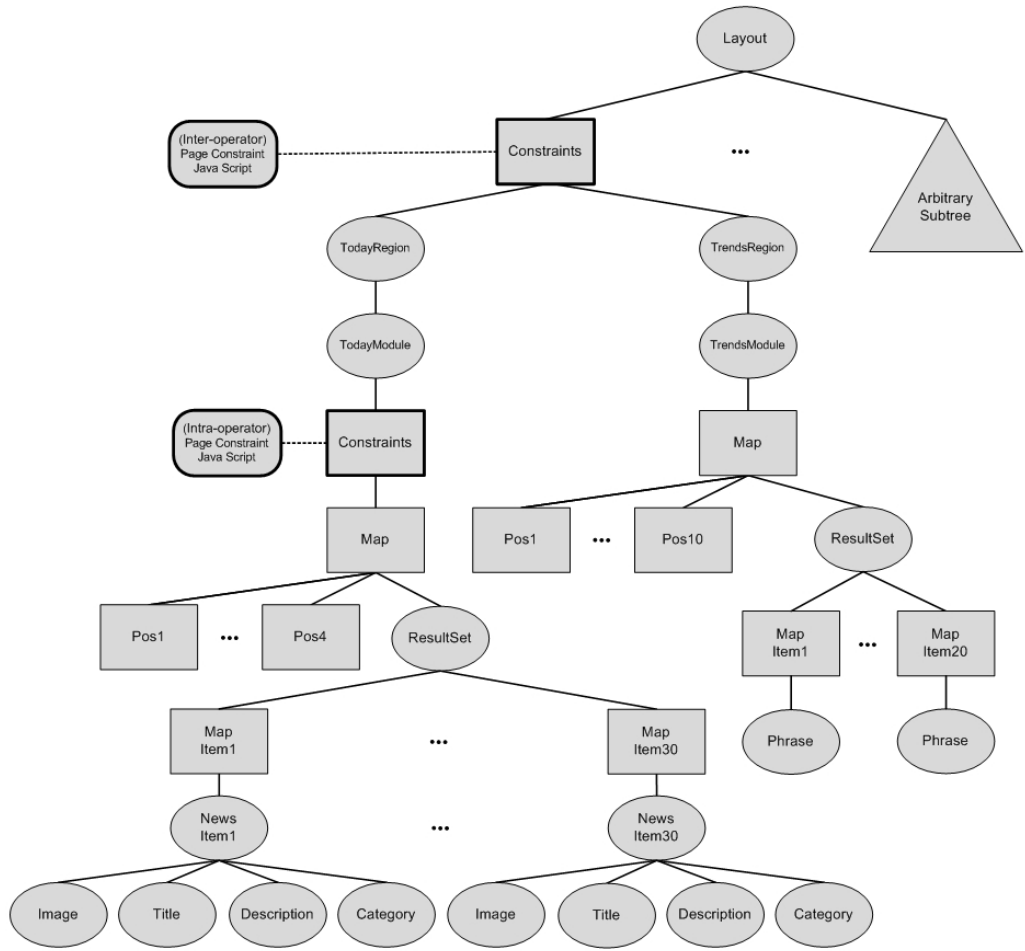

Fig. 7. A logical page model with embedded *constraints* operators.

To conclude, we note that the theoretical foundation of finding a feasible page instance given degrees of freedom and constraints is a generalization of the prototypical NP-hard *satisfiability* problem (SAT, [7]). Hence, there is no known worst-case polynomial-time algorithm for valid page generation. In practice, however, we typically expect many feasible page instances to exist, and that finding a valid solution will not be computationally hard. We assume that the difficult task would be choosing the "best" solution out of the numerous feasible ones (for more details see Section VI).

## VI. MODEL EXECUTION

Model *execution* is the process of resolving all FAME tags, which produces a plain XML that is ready for rendering (Section IV). This process translates to a sequence of operator evaluations.

Each *fetch, map* and *choice* operator in a page model defines an execution extension point. The model's designer must associate each such operator with an implementation plug-in, which we call *handler*. Handlers perform the actual fetching and optimization work – they retrieve data from sources, select news items, map content modules to page slots, etc. The FAME platform provides Java APIs for handler implementations. The built-in separation of concerns (Section II) allows isolating handler development from the XML model design and Javascript constraint programming[10]. The platform readily supports the sharing and reuse of handlers among multiple models[11].

The framework controls the order of handler execution. Individual handlers are unaware of the global execution flow – they only need to implement operator semantics. For example, a *map* handler must implement an API that assigns items to positions. It does not require any environmental information – e.g., the operator's location in the hierarchy, or whether its parameters are fixed or dynamically computed in the course of execution. The system mediates all communication between the handlers. This design principle is common to many data-flow architectures, e.g., database management systems [15].

The hierarchical page composition process confines the scope of each FAME operator to its XML subtree (Section IV). In other words, each operator must be evaluated before its output is required by its ancestors. This requirement is naturally fulfilled by the *bottom-up* execution order – i.e., each handler runs after all its dependencies are resolved, and produces a subtree that is free of FAME tags. The page model's XML (tree) structure therefore implicitly defines a workflow among its operators. For example, as depicted in Figures 6 and 7, a *map* handler that operates on a set of dynamic results is executed after the *fetch* handler for that source. Likewise, constraints on this subtree can be validated only after the items are fetched and mapped.

---

[10]Recall that these activities require profoundly different skill sets, and hence are separate by design.

[11]We use an open-source OSGi container implementation as a framework for flexible software component management.

In reality, the execution machinery is more involved, due to the need for constraint satisfaction. Since constraint operators can only return solutions (XML subtrees) that satisfy the constraint predicates, their descendant operators should be capable of producing more than one solution instance. In order to address this, FAME expects *map* and *choice* resolvers to expose their solution space for exploration at runtime, enabling the platform to modify subtrees whose top-level constraints fail.

### A. Handler Iterators

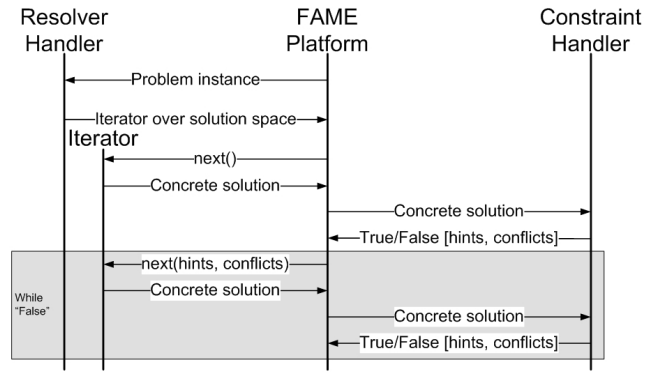

Fig. 8. Dynamic interaction between the handlers of a resolver operator and a constraint operator on top of it. The FAME platform orchestrates the execution order and propagates the constraint notifications to the resolver, indicating in which direction to explore the search space.

The API of *map* and *choice* resolvers has them exposing an *iterator* API – instead of returning a single solution (subtree), they expose their solution space for traversal via the *next()* method. Handlers of those resolvers are expected to traverse the space in *preference of instantiation* order, namely to first step through the solutions they most prefer to instantiate in the current page generation request. Typically those would be the best performing solutions, but at times they can be solutions whose performance needs to be assessed in explore/exploit experimentation schemes [8]. Thus, the iteration order of the same handlers may change between successive page requests of the same logical page.

The platform uses the iterators exposed by *map* and *choice* handlers to search for valid sub-trees rooted at each *constraints* operator. As FAME allows operators of all types to be composed hierarchically in the page model, constraint satisfaction translates to the orchestration of movements of embedded iterators. This is reminiscent of the manner by which search engines manipulate postings iterators over simple and complex search terms when evaluating queries over an inverted index. The literature offers a wealth of methods for enumerating large combinatorial spaces – from simple Backtracking, through Branch and Bound techniques, to heuristics such as Beam Search and Simulated Annealing, and more [16]. Most of these enumeration algorithms are suited for (or can be adapted to) the iterators' enumeration of solutions in decreasing preference of instantiation[12]. The platform instantiates the page once it finds a positioning of all iterators that satisfies all constraints.

Since operator handlers are unaware of the constraints imposed on the solutions they propose, their iteration order over a large solution space might require many steps till yielding a valid solution. To address this issue, we introduce a *constraint notification* mechanism. The idea is to indicate, in the constraint function, which predicate(s) could not be satisfied, in order to restrict the solution search when the embedded iterators get re-invoked. The feedback is propagated by the framework downstream, and passed to the descendant iterator's *next()* method as a parameter.

There are two types of constraint notifications – *hints* and *conflicts*. Hints are positive feedback – they indicate what must be done in order to obtain a valid solution. For example, a hint reported to a *map* operator might instruct to populate a particular position from a restricted set of items. Conflicts are negative feedback – i.e., what what must not be done. For instance, a conflict reported to a *choice* operator might say that its selected item may not be chosen in this context (typically due to an inter-operator constraint). Iterators receiving hints and conflicts are expected to fast-forward to the first-next solution in their iteration order that respects the notification.

Figure 8 illustrates the interaction between two FAME operators – a resolver $R$ embedded within a constraints operator $C$ – and the platform's mediation in this process. $R$'s handler gets invoked first (on some problem instance), and returns an iterator $\mathcal{I}$ over its solution space. The platform invokes *next()* on this iterator, retrieves a concrete solution, and passes it to $C$. The latter applies its embedded predicate function, which returns *false* with some hint or conflict notification. The platform then re-invokes $\mathcal{I}$.*next()*, which receives the notification as a parameter and outputs another potential solution. This process continues until $C$'s predicate holds for a solution returned by $\mathcal{I}$.

Since constraints result in some (partial) enumeration over the iterators of embedded resolvers, FAME model designers should minimize computations by placing constraint operators as deep as possible in the model's tree, to cover precisely their immediate scope (Section V). For example, an intra-operator constraint should optimally reside directly on top of the operator it validates. In mature database systems, such execution plan transformations are done automatically [15]. However, the potential of doing so in FAME is limited, since a general-purpose programming language is used for constraint functions and automatic determination of the constraints' minimal scope is difficult.

To summarize this discussion, we repeat our expectation of realistic page models having many valid instantiations (i.e. being very far from difficult SAT instances). Therefore, reasonable enumeration algorithms over the cross-product of resolvers' iterators, coupled with the mechanism of constraint notification and with each iterator's preference of instantiation order, typically converge quickly to valid and well-performing page instances.

## VII. HARNESSING USER FEEDBACK

Learning from real-time user feedback (*RTUF*) proved to be successful in multiple areas, such as content recommendation [2], [1], [3] and ranking of search results [12]. For example, in the news recommendation setting, multi-armed bandit explore-exploit schemes have been used to learn from user clicks on suggested stories and adapt to the audience interests online. In order to enable such optimization experiments, it is imperative for a FAME resolver (map/choice) to match user feedback to the decision that has selected or presented the content in some particular way.

Enabling this feedback loop requires close collaboration between the front-end, where user feedback is physically collected, and the resolver handler that conducts the optimization experiment. While the resolver is building its part of the logical page description, it must *instrument* the page for user feedback collection, i.e., embed into it all the context required for future feedback processing. The front-end, which renders this description, must respect the instrumentation directives, intercept the user actions (typing, clicks, mouse movements, etc.) that the resolver asked for, and propagate the collected data to the FAME back-end platform which serves as a dispatch mechanism that pushes the incoming feedback to the appropriate resolver.

Figure 9 depicts the end-to-end flow of information in a two-tiered system with a FAME back-end. For simplicity of presentation, we assume that the online learning algorithms run on the same platform that serves logical page descriptions. FAME therefore presents two façades – one for regular logical page requests, and another for RTUF processing. The former invokes the model execution machinery described in Section VI, which runs decision-making logic that *uses* the data learned from user feedback. The latter routes the feedback data stream to the learning algorithms that *update* the data repository in the background.

FAME optimization resolvers mirror the system's façades by having two interfaces – one for request handling, and another for RTUF processing. The framework offers a subscription API to plug into the feedback routing infrastructure, and an instrumentation API to decorate the operator's output with specifically designed tags. The instrumentation API enables a resolver to embed any instrumentation content in any format while keeping the actual information opaque to both the front-end and to the FAME platform. The platform is merely a pipe that delivers the feedback data, sent from the front-end, to its consuming plug-in.

There are three types of RTUF tags:

1) *Impression* tags record the optimization decisions. For example, a *map* handler may record the items that it has chosen for each position as well as the items it has rejected, in order to provide context for both positive and negative feedback.

---
[12]Some algorithms further require the iterators to return a fitness score at each step.

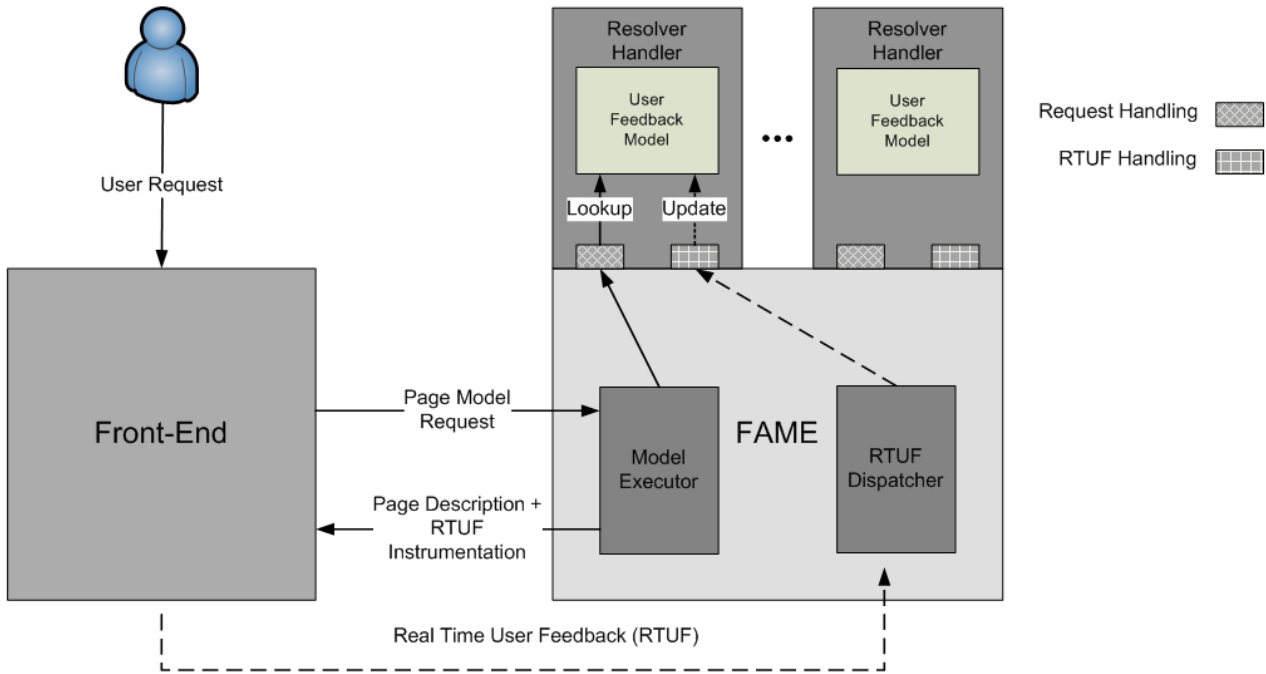

Fig. 9. An end-to-end architecture spanning a front-end and a FAME back-end. FAME has two façades: one for logical page generation, and another for RTUF processing.

2) *Item* tags designate the semantic items to be tracked (e.g., a link, an image or a module), in the context of a given impression.
3) *Action* tags designate the user actions to be tracked (e.g., clicks or touchscreen hovers) in the context of a given (impression, item) pair.

Figure 10 depicts an output subtree with instrumentation directives for the entire impression (news module) as well as a click over each news story. The front-end processes them as follows. First, when the physical page is served, it logs the impression tag with the page instance id. Next, upon each click-through, it logs the item tag in conjunction with the same id. These records are all streamed to the FAME back-end, where they are routed to the appropriate handler.

## VIII. Conclusions and Future Work

This work presented FAME, a framework for creating agile Web media experiences through algorithmic experimentation and optimization. The input to FAME are page instantiation requests expressed by logical page models, which define (1) dynamic data to be fetched, (2) degrees of freedom to algorithmically experiment with, subject to (3) editorial constraints, toward optimization of the page for user experience and business goals. During page instantiation, FAME's execution engine orchestrates various algorithmic plug-ins, which collectively morph the logical page model into a concrete page by resolving the model's degrees of freedom.

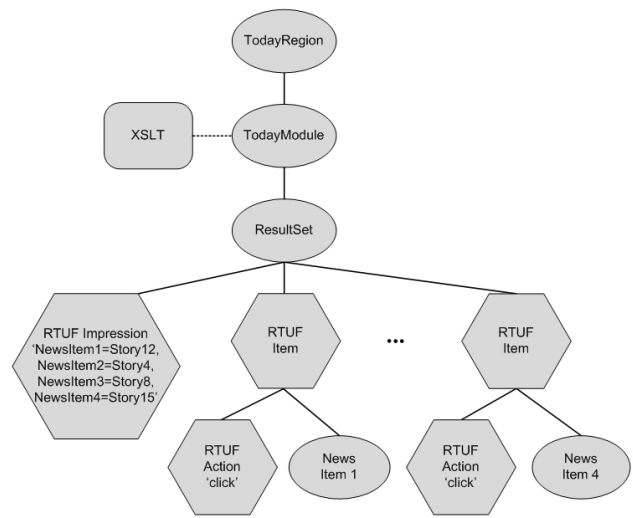

Fig. 10. A fragment of a logical page description instrumented with RTUF tags. The RTUF impression tag denotes the ids of the chosen news items, while the RTUF item tags encapsulate items for which user click feedback should be tracked.

FAME improves upon existing systems that perform multivariate optimization of Web pages in several aspects. It incorporates fetchers which import dynamic content into its page models in real time. It is further designed for hierarchical optimization of the page, meaning that it implicitly defines a workflow of decisions to be made and content to be fetched. In particular, this allows pages to be optimized given the attributes of the content that is available at page instantiation time. While FAME, like previous systems, defines multiple degrees of freedom to be resolved on its pages, it differs from those systems in that it uses a mechanism of constraints that avoids serving page instances with conflicting resolutions. Finally, FAME is an open platform, and defines interfaces and extension points that allow multiple optimization algorithms to be plugged in, each resolving specific portions of the page. Related work in Section III surveyed several reinforcement learning techniques which are natural plug-ins into FAME.

An important consequence of FAME's architecture is that the building blocks of its page models - hierarchical structure, fetchers, resolvers and constraints - naturally map to the responsibilities of different stakeholders in the online media serving pipeline. The architectural modularity enables a separation of concerns that allows people of different roles and skill sets – UED specialists, media editors, product owners and optimization experts – to work independently and then compose and reuse their artifacts in an agile manner.

Future work will shift the focus from the framework and platform aspect onto evaluations of complex optimization scenarios, where multiple resolvers implementing different learning algorithms are composed hierarchically and are orchestrated by the platform to reach well-performing whole-page operating points with respect to a spectrum of target functions.

## Acknowledgments

We thank Jon Bratseth, Jean-François Crespo, Kenneth Fox and Mike Wexler for many useful discussions.